\documentclass[11pt,twoside]{article}

\usepackage{asp2006}
\usepackage{epsf}
\usepackage{lscape}
\usepackage{graphicx}

\begin{document}
\title{
SPH Simulations of Accretion Flows onto the Supermassive Binary Black Holes
from the \\ Circumbinary Disks}   %%% Fill in title

\vspace{-0.5cm}
\author{Kimitake~Hayasaki$^1$ Shin~Mineshige$^1$ and Hiroshi~Sudou$^2$}
\affil{$^1$Yukawa Institute for Theoretical Physics,
           Oiwake-cho, Kitashirakawa, Sakyo-ku, Kyoto 606-8502 \\
       $^2$Faculty of Engineering, Gifu University, Gifu 501-1193}

\vspace{-0.1cm}
\begin{abstract} %%% Abstract to run on from here.
We investigate the accretion flows onto the supermassive binary black holes (SMBBHs) from the
circumbinary disk with the equal mass, eccentric binary on the subparsec scale,
using Smoothed Particle Hydrodynamics (SPH) code.
We find that the material can be supplied from circumbinary disk, 
which leads to the formation of two accretion disks around the SMBBHs.
The mass accretion rates significantly modulate with the binary orbital motion.
These could provide the observable diagnosis of the
existence of the supermassive binary black holes (e.g. OJ287) on the subparsec scale 
in merged galactic nuclei.
\end{abstract}

\vspace{-0.5cm}
\section{Introduction} 

The discovery of the tight correlation 
between black hole mass and the velocity dispersion 
of the bulge component of galaxies \citep{ferra, geb}
supports the scenario that the black holes have grown up in mass 
through the hierarchical galaxy mergers. 
This scenario inevitably would lead to 
the formation of the supermassive binary black holes (SMBBHs)
during the course of galaxy mergers.
However, There has so far been little observational
direct evidence of the SMBBHs on the subparsec scale \citep{sudou}.

If there is the gas orbiting around the SMBBHs on the subparsec scale,
one will be able to observe a signal arising from
the interaction between the binary and its surrounding gas
(i.e. a circumbinary disk).
Therefore, we study the accretion onto the SMBBHs from the circumbinary disks
, performing the 3D Smoothed Particles Hydrodynamic (SPH) simulations.

\vspace{-0.5cm}
\section{Accretion flows onto the supermassive binary black holes} 

Our simulations were performed by using the same 3D SPH code as 
\cite{ba,haya}.
In order to investigate the accretion flows from the circumbinary disk,
we run the simulation with 
the Shakura-Sunyaev viscosity parameter
$\alpha_{\rm{SS}}=0.1$ throughout the circumbinary disk
which is coplanar with the orbital plane.
The orbital period $P_{\rm{orb}}$ is
about $296\,\rm{yr}$, the eccentricity $e$ is 0.5, and 
the total black holes masses $M_{\rm{bh}}=10^{8}M_{\odot}$ with the mass ratio $q=1.0$.
The black holes are modeled by a couple of sink particles with
the accretion radius
$r_{\rm{in}}=2.0\times10^{-2}a$, where $a=0.1\,\rm{pc}$ is the semi-major axis
of the binary. 
The isothermal equation of state is adopted with $T_{\rm{eff}}=10^{4}\,K$,
which approximately equals to 
the typical effective temperature of a standard disk 
at the $r_{\rm{in}}=2.0\times10^{-2}a$
around a single black hole with $10^{8}M_{\odot}$.

Fig.~1 shows the accretion flow onto the SMBBHs (the left panel) and the orbital phase
dependence of the mass-accretion rate onto the black holes (right panel), respectively.
From the left panel of Fig.~1, 
we note that the mass can inflow via two points at the inner edge of the circumbinary disk
onto the black holes, and then the accretion disks are formed around 
each of black holes.
In the right panel of Fig.1, to reduce the fluctuation noise,
the data are folded on the orbital period
over $40\le{t}\le60$.
From the figure, we note that the mass accretion rate strongly depends on
the binary orbital phase.

%%%%%%%%%%%%%%%%%%%%%%%%%%%%%%%%%%%%%%%%%
%%%%%%%%%%%%%%%% FIGURES %%%%%%%%%%%%%%%%
%%%%%%%%%%%%%%%%%%%%%%%%%%%%%%%%%%%%%%%%%
\begin{figure}[!ht]
\begin{center}
\includegraphics[width=3.8cm]{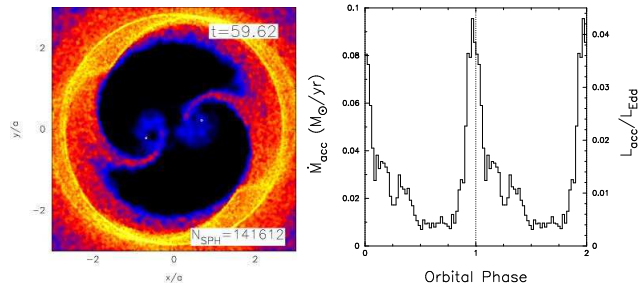}
\includegraphics[width=4.4cm]{kimitake_hayasaki_fig2.eps}
\end{center}
\vspace{-0.5cm}
\caption{
Snapshot of accretion flow from the circumbinary disk 
onto the supermassive BBHs (the left panel).
and orbital-phase dependence of mass-accretion rate (the right panel). 
In the left panel, 
the origin is set on the center of mass of the binary and all panels
are shown in a binary rotation frame.
The surface density is shown in a range of three orders
of magnitude in the logarithmic scale.
The solid circles denote a couple of black holes with the
accretion radius $r_{\rm{in}}=2.0\times10^{-2}a$.
Annotated in the panel are the time in units of $P_{\rm{orb}}$
and the number of SPH particles $N_{\rm{SPH}}$.
In the right panel, 
the right axis shows the bolometric luminosity corresponding to the
mass-accretion rate with the energy conversion efficiency, $\eta=0.1$,
normalized by the Eddington luminosity with total black hole mass
$M_{\rm{bh}}=10^{8}M_{\odot}$.
}
\label{fig1}
\end{figure}

\vspace{-1.0cm}
\section{Summary}
We have carried out the SPH simulations
of accretion flows from circumbinary disks onto the SMBBHs
on the subparsec scale.
We find that
the two accretion disks are formed around the SMBBHs.
The mass accretion rate significantly varies
with an orbital motion because of
the periodic variations of the binary potential.
This could provide the observable diagnosis
for the presence of the SMBBHs (e.g. OJ287) in three-disk systems
at the galactic center.
 
\acknowledgements %%% Text of acknowledgements runs on after this command.
The simulations reported here were performed using the facility
of the Centre for Astrophysics \& Supercomputing at
Swinburne University of Technology, Australia and 
of YITP in Kyoto University.
This work has been supported
by the Grants-in-Aid of the Ministry
of Education, Science, Culture, and Sport and Technology (MEXT;
14079205 K.H. \& S.M., 16340057 S.M.).

%%% THE BIBLIOGRAPHY
%%%
%%% CONSULT SECTION 3 OF "INSTRUCTIONS FOR AUTHORS" FOR HOW TO USE NATBIB.
%%% AUTHORS ARE ENCOURAGED TO USE EITHER THE "THEBIBLIOGRAPY" ENVIRONMENT
%%% BY UNCOMMENTING (DELETING THE "%" SYMBOL) THE COMMANDS BELOW, OR BY
%%% USING THE BIBTEX ENVIRONMENT. TO FIND OUT WHICH IS APPLICABLE TO YOUR
%%% CONTRIBUTION, CONSULT THE VOLUME EDITORS FOR YOUR PROCEEDINGS.
%%%

\end{document}